\documentclass[sigconf]{acmart}
\AtBeginDocument{%
  }

\copyrightyear{2027}
\acmYear{2027}
\setcopyright{cc}
\setcctype{by}
\acmConference[WSDM '27]{Proceedings of the Twentieth ACM International Conference on Web Search and Data Mining}{February 15--19, 2027}{Hong Kong, Hong Kong}
\acmBooktitle{Proceedings of the Twentieth ACM International Conference on Web Search and Data Mining (WSDM '27), February 15--19, 2027, Hong Kong, Hong Kong}
\acmDOI{10.1145/3862322.3863794}
\acmISBN{979-8-4007-3117-4/2027/02}

\usepackage{booktabs}
\usepackage{multirow}
\usepackage{makecell}
\usepackage{tabularx}
\usepackage{array}
\newcolumntype{Y}{>{\raggedright\arraybackslash}X}

\usepackage{colortbl}
\definecolor{cogrecgray}{gray}{0.93}

\begin{document}
\title{CogRec: Structure-Cognitive Fast-and-Slow Reasoning for Generative Recommendation}

\makeatletter

\newcommand{\cogrec@emails}{}

\newcommand{\cogrec@queueemail}[2]{%
  \g@addto@macro\cogrec@emails{%
    \par\bgroup
    \mathchardef\UrlBreakPenalty=10000
    \href{mailto:#2}{#2}\egroup
  }%
}

\newcommand{\cogrec@patcherror}{%
  \PackageError{cogrec-authors}{Author layout patch failed}%
    {This author layout was checked with acmart v2.18.}%
}

\patchcmd{\@mkauthors@iii}
  {\def\affiliation}
  {\let\email\cogrec@queueemail\def\affiliation}
  {}{\cogrec@patcherror}

\patchcmd{\@typeset@author@bx}
  {\@currentaffiliation}
  {\@currentaffiliation\cogrec@emails}
  {}{\cogrec@patcherror}

\apptocmd{\@typeset@author@bx}
  {\gdef\cogrec@emails{}}
  {}{\cogrec@patcherror}

\newcommand{\CogRecAff}[1]{%
  \affiliation{%
    \institution{#1}%
    \city{Beijing}%
    \country{China}}
}

\newcommand{\CogRecFirstAff}[1]{%
  \affiliation[obeypunctuation=true]{%
    \institution{#1}%
    \begingroup
      \let\@ACM@addtoaddress\@gobble
      \city{Beijing}\country{China}%
    \endgroup
  }%
}

\makeatother

\settopmatter{authorsperrow=3}


\author{Xiang Liu}
\authornote{These authors contributed equally to this work.}
\CogRecFirstAff{Institute of Information Engineering, Chinese Academy of Sciences}
\CogRecAff{School of Cyber Security, University of Chinese Academy of Sciences}
\email{liuxiang@iie.ac.cn}

\author{Jingsong Su}
\authornotemark[1]
\CogRecAff{School of Artificial Intelligence, Beijing Normal University}
\email{sujingsong@mail.bnu.edu.cn}

\author{Shuqi Zhao}
\email{zhaoshuqi@iie.ac.cn}

\author{Pengbo Mo}
\email{mopengbo@iie.ac.cn}

\CogRecAff{Institute of Information Engineering, Chinese Academy of Sciences}


\author{Yiming Qiu}
\CogRecAff{JD.com}
\email{qiuyiming3@jd.com}

\author{Huimu Wang}
\CogRecAff{Institute of Automation, Chinese Academy of Sciences}
\email{huimu.wang@ia.ac.cn}

\author{Mingming Li}
\authornote{Corresponding author.}
\CogRecAff{Institute of Information Engineering, Chinese Academy of Sciences}
\email{limingming@iie.ac.cn}


\author{Jiao Dai}
\CogRecAff{Institute of Information Engineering, Chinese Academy of Sciences}
\email{daijiao@iie.ac.cn}

\author{Jizhong Han}
\CogRecAff{Institute of Information Engineering, Chinese Academy of Sciences}
\email{hanjizhong@iie.ac.cn}

\author{Songlin Hu}
\CogRecFirstAff{Institute of Information Engineering, Chinese Academy of Sciences}
\CogRecAff{School of Cyber Security, University of Chinese Academy of Sciences}
\email{husonglin@iie.ac.cn}


\renewcommand{\shortauthors}{Xiang Liu et al.}


\begin{abstract}
Generative recommendation reformulates next-item prediction as identifier sequence generation, yet existing methods mainly treat the generated identifiers as target sequences to be memorized, leaving their hierarchical and relational structure underused for explicit reasoning.
We propose CogRec, a structure-cognitive fast-and-slow reasoning framework that grounds intermediate reasoning in the same SID topology used for target generation. CogRec augments the vertical SID hierarchy with intra-layer semantic graphs and introduces SID Routing to represent recommendation reasoning through layer-wise \textsc{Match}, \textsc{LateralJump}, and \textsc{Explore} operations. These operations provide a structural interpretation of fast localization and slow navigation. A supervised multi-stage pipeline aligns newly introduced SID tokens, establishes direct generation, and trains natural-language and SID Routing branches from a shared checkpoint.
Experiments on three Amazon Reviews 2014 categories show dataset-dependent effects of SID Routing. Routing improves Hit@10 and NDCG@10 over its direct-generation counterpart on Beauty and Sports, while reducing both metrics on Toys. These results characterize the benefits and dataset-dependent behavior of SID-grounded intermediate reasoning for generative recommendation. Code is available at \url{https://github.com/caskcsg/CogRec}.
\end{abstract}





\maketitle

\section{Introduction}
Generative recommendation reformulates next-item prediction as identifier sequence generation. A representative line of work encodes items as hierarchical Semantic IDs (SIDs), enabling the use of language models for item generation~\cite{tay2022dsi,rajput2023tiger,liu2025onerec}. Compared with atomic item identifiers, SIDs provide a structured output space and make recommendation compatible with sequence-to-sequence modeling.

Current SID-based methods, however, primarily treat SIDs as target sequences to be memorized. The hierarchical structure offers coarse-to-fine semantic organization, but is rarely used as an explicit reasoning substrate. Lateral relations among semantically related codes at the same layer are typically ignored. As a result, the model may generate valid identifiers, but the intermediate process from historical items to the target remains unstructured.

Reasoning-enhanced recommendation methods attempt to address this limitation. ReaRec performs implicit multi-step reasoning over hidden representations of a sequential recommender~\cite{tang2025rearec}. OneRec-Think generates an explicit textual rationale before the target identifier~\cite{liu2025onerecthink}. These studies show that additional computation can improve recommendation when user intent is hard to resolve. However, latent hidden-state iteration and natural-language rationales are not directly coupled with the discrete SID decoding space. Textual reasoning operates in language tokens, but the final output must be a valid hierarchical SID sequence. This representational difference motivates studying intermediate traces expressed directly in the SID space.
Unlike prior work that focuses on improving identifier construction or semantic alignment, we instead keep the SID construction fixed and aim to exploit the resulting topology itself as an explicit reasoning space for structured navigation.

Motivated by this representational difference, we propose CogRec, a structure-grounded fast-and-slow reasoning framework for generative recommendation. CogRec augments the SID hierarchy with intra-layer semantic graphs and defines SID Routing through layer-wise \textsc{Match}, \textsc{LateralJump}, and \textsc{Explore} operations. These operations express structural relations between historical items and target identifiers. Exact matching corresponds to fast localization, while lateral and exploratory operations provide a structural interpretation of slow navigation. The intermediate traces are expressed in the SID space used for final target generation, explicitly coupling the reasoning representation with the prediction space.

A supervised multi-stage pipeline aligns SID token embeddings, establishes direct generation, and trains natural-language and SID Routing branches from a shared checkpoint under identical SID mapping and decoding constraints. This controlled setup isolates the effect of reasoning representation from confounding factors of identifier construction.


We evaluate CogRec on three public sequential recommendation benchmarks. Under the reported settings, SID Routing improves Hit@10 and NDCG@10 over CogRec-Direct on Beauty and Sports, while reducing both metrics on Toys. Comparisons with natural-language reasoning and difficulty-stratified analyses further reveal that the effects of intermediate reasoning vary across datasets and structural conditions.





The contributions of this paper are summarized as follows:
\begin{itemize}
    \item We construct a SID topology that augments the output hierarchy with intra-layer semantic graphs, enabling intermediate reasoning to be expressed in the same structured space as target generation.

    \item We define SID Routing through layer-wise \textsc{Match}, \textsc{LateralJump}, and \textsc{Explore} operations, providing an explicit structural interpretation of fast-and-slow reasoning without introducing separate neural modules.

    \item We establish a multi-stage pipeline that first learns direct SID generation and then trains natural-language and SID Routing branches from a shared checkpoint, holding the SID mapping and initialization fixed for the reasoning-branch comparison.

    \item We evaluate the resulting reasoning formats on three public benchmarks and analyze their behavior across structural difficulty groups. The results document both improvements and degradations, highlighting the dataset-dependent effects of SID Routing.
\end{itemize}

\section{Related Work}

\subsection{Traditional Sequential Recommendation}
Traditional sequential recommendation methods model user behavior sequences or user-item interaction graphs for next-item prediction.
  Representative methods include recurrent architectures such as GRU4Rec~\cite{hidasi2016gru4rec}, self-attention models such as SASRec~\cite{kang2018sasrec}, bidirectional encoders such as BERT4Rec~\cite{sun2019bert4rec}, hierarchical gating models such as HGN~\cite{ma2019hgn}, and large-scale sequential transducers such as HSTU~\cite{zhai2024hstu}. Graph-based collaborative filtering methods such as NGCF~\cite{wang2019ngcf} and LightGCN~\cite{he2020lightgcn} provide complementary user-item relation modeling. These approaches operate on atomic item identifiers or learned embeddings rather than generating structured Semantic IDs.


\subsection{Generative Recommendation}
Generative retrieval reformulates retrieval as identifier generation, as demonstrated by DSI~\cite{tay2022dsi}, while TIGER introduces hierarchical Semantic IDs into sequential recommendation~\cite{rajput2023tiger}. Discrete identifier construction builds on vector quantization and product quantization techniques~\cite{oord2017vqvae,jegou2011product}. OneRec unifies retrieval and ranking within a generative recommendation framework~\cite{liu2025onerec}. More recent approaches improve Semantic-ID learning from different angles: LETTER combines semantic, collaborative, and code-diversity objectives for item tokenization~\cite{wang2025letter}; EAGER integrates behavioral and semantic streams~\cite{zhang2025eager}; GRAM incorporates hierarchical item relations via multi-granular fusion~\cite{lee2025gram}; LOHRec exploits order and hierarchy in lightweight generative recommendation~\cite{xie2025lohrec}; DIGER makes SID learning differentiable with respect to the recommendation objective~\cite{fu2026diger}; and CapsID studies soft-routed variable-length Semantic IDs~\cite{cheng2026capsid}. These methods improve identifier construction or semantic alignment, but generally do not formulate the SID topology itself as an explicit layer-wise reasoning path. Broader language-model-based recommendation approaches, such as P5~\cite{geng2022p5}, represent recommendation tasks in a unified token space, and recent surveys summarize this growing intersection~\cite{wu2024llmrecsurvey}.

\begin{figure*}[tp]
    \centering
    \includegraphics[width=0.98\textwidth]{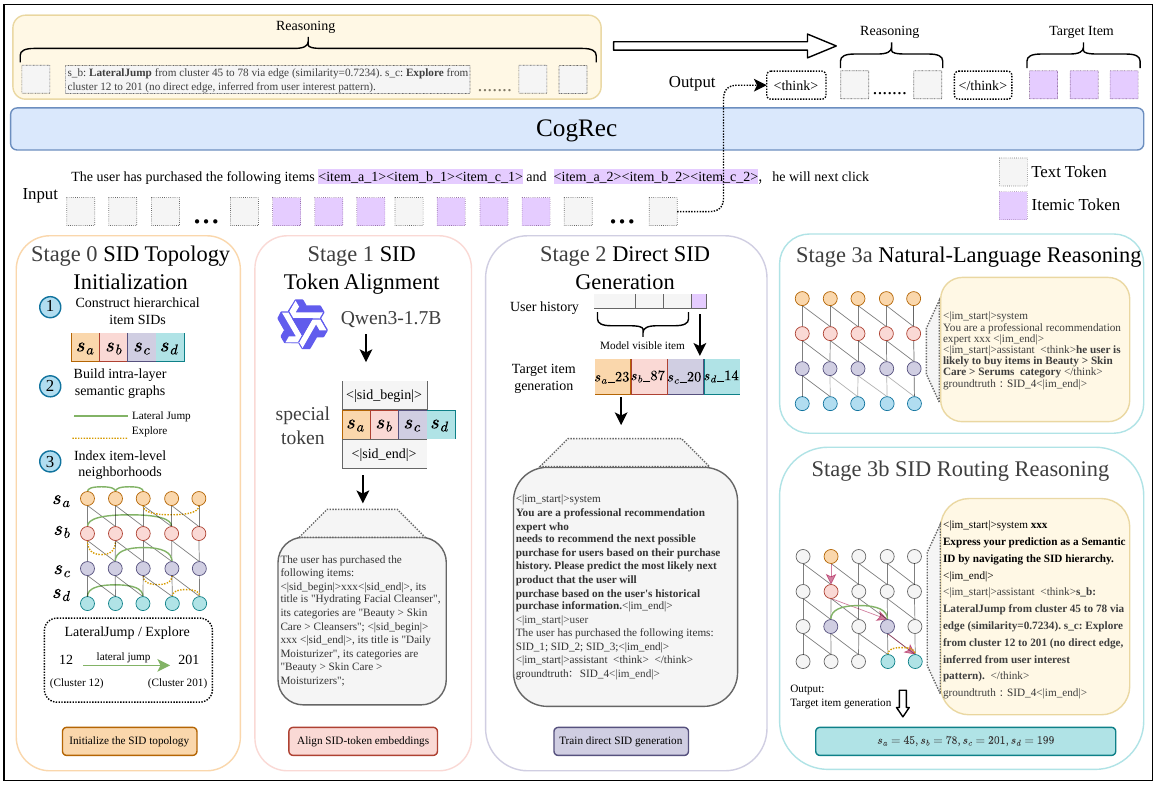}
    \caption{
        Overview of the CogRec framework. Stage 0 constructs the SID topology. Stage 1 aligns SID tokens. Stage 2 trains direct generation. Stage 3a and Stage 3b learn natural-language reasoning and SID Routing, respectively. }
    \Description{
        A left-to-right pipeline diagram of CogRec. Stage 0 constructs four-layer Semantic IDs, intra-layer semantic graphs, and item-level neighborhoods. Stage 1 aligns the newly added SID tokens with item metadata. Stage 2 trains direct target-SID generation. Two branches start from the Stage 2 checkpoint: Stage 3a generates a natural-language rationale before the target SID, and Stage 3b generates layer-wise SID Routing operations before the target SID.
        }
    \label{fig:cogrec_overview}
\end{figure*}

\subsection{Reasoning-Enhanced Recommendation}
Reasoning-enhanced recommendation includes latent iterative computation in conventional sequential recommendation and explicit or implicit reasoning in LLM-based generative recommendation. ReaRec augments sequential backbones with inference-time implicit multi-step reasoning over hidden representations~\cite{tang2025rearec}. In generative recommendation, OneRec-Think generates explicit reasoning before item-identifier generation~\cite{liu2025onerecthink}; PauseRec adds implicit reasoning without explicit rationale supervision~\cite{zhang2026pauserec}; TwiSTAR studies adaptive fast-and-slow allocation~\cite{cao2026twistar}; and SIDReasoner strengthens SID-language alignment with outcome-driven optimization~\cite{he2026sidreasoner}. Unlike these approaches, CogRec derives deterministic, layer-wise supervision directly from the SID topology and represents reasoning with \textsc{Match}, \textsc{LateralJump}, and \textsc{Explore} operations.

\subsection{Fast-and-Slow Reasoning}
Dual-process theory distinguishes fast intuitive judgment from slow deliberative reasoning~\cite{kahneman2011thinking}. Related AI techniques include chain-of-thought~\cite{wei2022cot}, self-consistency~\cite{wang2023selfconsistency}, Tree of Thoughts~\cite{yao2023tot}, ReAct~\cite{yao2023react}, least-to-most prompting~\cite{zhou2023leasttomost}, and adaptive reasoning studies~\cite{pan2025dynathink,chung2025thinker}. CogRec instantiates this perspective in the SID space: fast reasoning corresponds to vertical SID matching, while slow reasoning corresponds to lateral jumping and exploration over a structure-cognitive SID topology.

Different from methods that improve SID construction or semantic alignment, CogRec treats the fixed SID topology as an explicit reasoning substrate rather than a pure target space. Unlike reasoning-enhanced approaches that generate natural-language rationales, CogRec grounds all reasoning operations directly in the SID topology used for target generation, providing a structure-grounded alternative that is orthogonal to existing identifier construction methods.

\section{Method}
\label{sec:method}

\subsection{Problem Formulation}
\label{sec:problem_formulation}

Let $\mathcal{U}$ and $\mathcal{I}$ denote the user and item sets. For each user $u \in \mathcal{U}$, the historical interaction sequence is $H_u=[i_1,i_2,\ldots,i_t]$, where $i_j \in \mathcal{I}$. The next-item recommendation task is to predict the target item $i^\star=i_{t+1}$ conditioned on $H_u$.

In SID-based generative recommendation, each item $i$ is represented by a hierarchical Semantic ID $z_i=[z_i^1,z_i^2,\ldots,z_i^L]$, where $L$ is the number of layers and $z_i^l$ is the discrete code at layer $l$. We set $L=4$ and each layer contains 256 codes. The SID is serialized as:
\begin{equation}
    \langle\texttt{sid\_begin}\rangle
    \langle\texttt{s\_a}\rangle
    \langle\texttt{s\_b}\rangle
    \langle\texttt{s\_c}\rangle
    \langle\texttt{s\_d}\rangle
    \langle\texttt{sid\_end}\rangle.
\end{equation}

CogRec supports two generation formats. Direct SID generation models $p_\theta(z_{i^\star}\mid H_u)$ and predicts the target identifier without an intermediate reasoning trace. SID Routing additionally generates a route $R_u$:
\begin{equation}
    p_\theta(R_u,z_{i^\star}\mid H_u)
    =
    p_\theta(R_u\mid H_u)
    p_\theta(z_{i^\star}\mid H_u,R_u).
\end{equation}
Both formats share the same item-to-SID mapping, candidate item space, and decoding constraint.

\subsection{Overview}
CogRec grounds recommendation reasoning in the Semantic ID space used for target generation. As shown in Figure~\ref{fig:cogrec_overview}, its workflow comprises four stages: SID topology construction (Stage 0), SID-token alignment (Stage 1), direct SID generation (Stage 2), and reasoning training (Stage 3). Stage 2 provides a shared checkpoint for two alternative Stage 3 branches: natural-language reasoning (Stage 3a) and structure-grounded SID Routing (Stage 3b).

\subsection{Structure-Cognitive SID Space}
\label{sec:sid_space}

Standard SID-based methods use hierarchical identifiers mainly as output sequences. CogRec constructs a navigable topology that includes vertical hierarchy, intra-layer semantic relations, and item-level neighborhoods.

\textbf{Hierarchical Semantic IDs.}
Let $x_i$ denote the metadata text of item $i$, and let $e_i=f_{\mathrm{enc}}(x_i)$ be its normalized semantic embedding. Residual K-means quantization with layer-specific codebooks $\mathcal{C}^{(l)}=\{c_1^{(l)},\ldots,c_K^{(l)}\}$ produces the codes. Starting from $r_i^{(0)}=e_i$, the code and residual at layer $l$ are
\begin{align}
    z_i^l
    &=
    \arg\min_{k\in\{1,\ldots,K\}}
    \left\|
    r_i^{(l-1)}-c_k^{(l)}
    \right\|_2^2,\\
    r_i^{(l)}
    &=
    r_i^{(l-1)}-c_{z_i^l}^{(l)}.
\end{align}
The first layer quantizes the item embedding itself. Each subsequent layer quantizes the residual left after subtracting the centroids selected at earlier layers. When two items share a SID prefix of length $l$, they have been assigned the same centroid at each of the first $l$ quantization layers. A longer shared prefix therefore localizes the two items to a finer region of the SID hierarchy.

\textbf{Intra-layer semantic graphs.}
For each layer $l$, we construct a graph $G^{(l)}=(V^{(l)},E^{(l)})$, where each node is a centroid. For centroids $c_a^{(l)}$ and $c_b^{(l)}$, their relation score is
\begin{equation}
    q_{ab}^{(l)}
    =
    \frac{
    \langle c_a^{(l)},c_b^{(l)}\rangle
    }{
    \|c_a^{(l)}\|_2\|c_b^{(l)}\|_2
    }.
\end{equation}
For each node $a$, we retain edges to the top-$k$ centroid neighbors whose similarity is at least $\tau$:
\begin{equation}
    E^{(l)}
    =
    \left\{
    (a,b)
    \,\middle|\,
    b\in\mathcal{N}_k^{(l)}(a),
    \ q_{ab}^{(l)}\geq\tau
    \right\}.
\end{equation}
These edges preserve semantic proximity between codes at the same abstraction level.

\textbf{Item-level neighborhoods.}
Code-level graphs operate on centroids and may lose fine-grained relations between individual items. We additionally construct a hierarchical navigable small-world index over item embeddings, which retains local semantic neighborhoods that complement the centroid graphs during offline construction and diagnosis.

Together, the hierarchy and lateral relations transform the SID space from a pure output tree into a topology supporting both fast vertical localization and slower lateral navigation.

\begin{figure}[t]
    \centering
    \includegraphics[width=\linewidth]{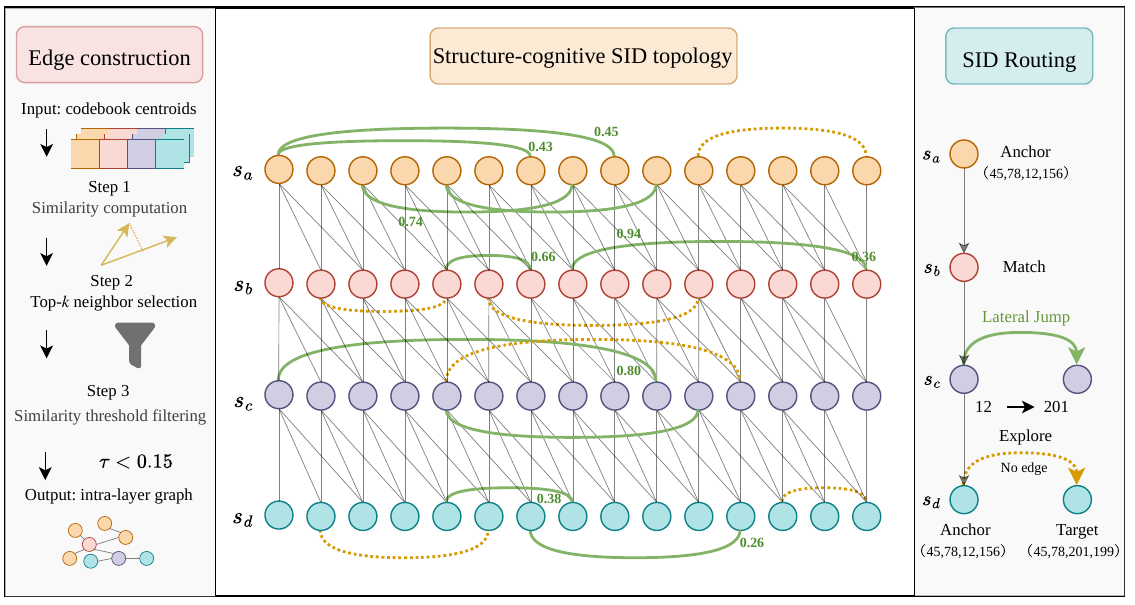}
    \caption{
    Structure-cognitive SID space and SID Routing. The intra-layer graph is constructed by selecting top-$k$ centroid neighbors and filtering candidate edges according to a similarity threshold $\tau$. Given a historical anchor and a target item during offline trace construction, SID Routing derives a layer-wise route consisting of \textsc{Match}, \textsc{LateralJump}, and \textsc{Explore} operations.
    }
    \Description{
    A three-part diagram. The left panel computes centroid similarities, selects top-k neighbors, and filters candidate edges according to a similarity threshold. The center panel shows a hierarchical SID space augmented with retained lateral semantic edges and item-level neighborhoods. The right panel illustrates an anchor-to-target route containing Match, LateralJump, and Explore operations.
    }
    \label{fig:sid_space_routing}
\end{figure}

\subsection{Fast-and-Slow SID Routing}
\label{sec:sid_routing}

SID Routing converts recommendation reasoning into layer-wise operations over the SID topology. The right panel of Figure~\ref{fig:sid_space_routing} illustrates this construction for one historical anchor and target item, showing how the relation between their codes at each layer determines the corresponding routing operation. The target-dependent definitions in this subsection are used only for offline supervision construction and diagnostics. During inference, the model receives only the user history; the target, anchor relations, and reference route are not provided.

\textbf{Historical anchor selection.}
For a historical item $h\in H_u$ and target $i^\star$, their prefix-matching depth is
\begin{equation}
    m(h,i^\star)
    =
    \max
    \left\{
    \rho\in\{0,\ldots,L\}
    \,\middle|\,
    z_h^l=z_{i^\star}^l,\ \forall l\leq\rho
    \right\}.
\end{equation}
The maximum depth over the history is $M_u=\max_{h\in H_u}m(h,i^\star)$, and the corresponding item is selected as the anchor:
\begin{equation}
    a_u=\arg\max_{h\in H_u}m(h,i^\star).
\end{equation}

If no historical item matches the target at the first layer, the last item in the sequence is used as a deterministic fallback anchor. The same anchor is used at all four layers, so a later layer can still receive a \textsc{Match} even after an earlier mismatch has ended the shared prefix.

\textbf{Layer-wise routing operations.}
For each layer $l$, CogRec compares the anchor code $z_{a_u}^l$ and the target code $z_{i^\star}^l$:
\begin{equation}
r_l =
\begin{cases}
\textsc{Match},
&
z_{a_u}^l=z_{i^\star}^l,
\\
\textsc{LateralJump},
&
(z_{a_u}^l,z_{i^\star}^l)\in E^{(l)},
\\
\textsc{Explore},
&
\text{otherwise}.
\end{cases}
\end{equation}
A \textsc{Match} indicates identical codes. A \textsc{LateralJump} indicates an explicit intra-layer edge. An \textsc{Explore} is assigned when neither exact match nor a graph edge is observed. The complete route is $R_u=[r_1,r_2,\ldots,r_L]$. For supervised training, only non-trivial operations are verbalized in the \texttt{<think>} segment; \textsc{Match} operations may be omitted.

\textbf{Operational fast-and-slow computation.}
We assign an auxiliary structural cost to each operation:
\begin{equation}
c(r_l)=
\begin{cases}
0, & r_l=\textsc{Match},\\
1, & r_l=\textsc{LateralJump},\\
2, & r_l=\textsc{Explore}.
\end{cases}
\end{equation}


The total cost and number of non-\textsc{Match} steps are
\begin{equation}
    C_u=\sum_{l=1}^{L}c(r_l),
    \quad
    s_u=\sum_{l=1}^{L}\mathbb{I}[r_l\neq\textsc{Match}].
\end{equation}

A route with $C_u=0$ contains only \textsc{Match} operations, whereas $C_u>0$ indicates at least one non-\textsc{Match} operation. For example, a route with a single \textsc{LateralJump} and a route with a single \textsc{Explore}, with all other layers matched in both cases, each have $s_u=1$ but incur costs of $C_u=1$ and $C_u=2$, respectively. The values $0,1,2$ define an ordinal diagnostic convention and are not learned costs or measured latency.

\subsection{Structural Difficulty and Routing Complexity}
\label{sec:difficulty_step}
The same topology used to construct routing traces provides two complementary diagnostics: coarse structural difficulty and fine-grained routing complexity.

\textbf{Graph-based proximity.}
Let $G^{(1)}$ denote the graph at the first and coarsest layer. For each user-target pair, the minimum history-to-target graph distance is
\begin{equation}
    D_u
    =
    \min_{h\in H_u}
    \operatorname{dist}_{G^{(1)}}
    \left(
    z_h^1,z_{i^\star}^1
    \right).
\end{equation}
We cap the breadth-first search in the first-layer graph at $D_{\max}=4$ hops. If the target code cannot be reached within this limit from any first-layer code in the user's history, we assign $D_u=D_{\max}+1$ to record that no path was found within four hops.

\textbf{Difficulty groups.}
Using the maximum prefix depth $M_u$ and graph distance $D_u$, we define
\begin{equation}
\delta_u =
\begin{cases}
\textsc{Easy},
&
M_u\geq2,
\\
\textsc{Medium},
&
M_u<2
\ \land\
(M_u\geq1\ \lor\ D_u\leq2),
\\
\textsc{Hard},
&
\text{otherwise}.
\end{cases}
\end{equation}
Easy instances have a historical item sharing at least the first two SID layers with the target. Medium instances lack a sufficiently long prefix but preserve first-layer agreement or short-range graph proximity. Hard instances have neither. These labels are used for analysis and diagnosis rather than as training targets.

The difficulty label $\delta_u$ describes coarse proximity, whereas $s_u$ and $C_u$ measure the actual layer-wise complexity of the constructed route. For example, if the anchor and target match at the first two layers and differ at the last two, the instance is \textsc{Easy}, but its route still contains two non-\textsc{Match} steps. Section~\ref{sec:difficulty_analysis} uses $\delta_u$ to group instances when comparing recommendation performance, and Section~\ref{sec:routing_behavior} reports the distribution of $s_u$.
\begin{table*}[tp]
    \centering
    \scriptsize
    \setlength{\tabcolsep}{2.0pt}
    \renewcommand{\arraystretch}{1.10}
    \caption{
        Performance comparison across three categories of recommendation methods.
        Bold and underlined values indicate the best and second-best results, respectively.
    }
    \label{tab:overall_results}

    \resizebox{\textwidth}{!}{%
    \begin{tabular}{
        ll
        *{5}{c}
        @{\hspace{4pt}}
        *{4}{c}
        >{\columncolor{cogrecgray}}c
        @{\hspace{20pt}}
        *{4}{c}
        >{\columncolor{cogrecgray}}c
    }
        \toprule

        \textbf{Dataset}
        &
        \textbf{Metric}
        &
        \multicolumn{5}{c}{\textbf{Traditional Methods}}
        &
        \multicolumn{5}{c}{\textbf{Generative Methods}}
        &
        \multicolumn{5}{c}{\textbf{Reasoning Methods}}
        \\
        \cmidrule(lr){3-7}
        \cmidrule(lr){8-12}
        \cmidrule(lr){13-17}

        &
        &
        \textbf{BERT4Rec}
        &
        \textbf{HGN}
        &
        \textbf{GRU4Rec}
        &
        \textbf{SASRec}
        &
        \textbf{HSTU}
        &
        \textbf{TIGER}
        &
        \textbf{LETTER}
        &
        \textbf{EAGER}
        &
        \textbf{\makecell{OneRec-Think\\Direct}}
        &
        \textbf{\makecell{CogRec\\Direct}}
        &
        \textbf{ReaRec}
        &
        \textbf{PauseRec}
        &
        \textbf{TwiSTAR}
        &
        \textbf{\makecell{OneRec-\\Think}}
        &
        \textbf{\makecell{CogRec\\Routing}}
        \\
        \midrule

        \multirow{4}{*}{\textbf{Beauty}}
        & Hit@5
        & 0.0232
        & 0.0319
        & 0.0395
        & 0.0402
        & 0.0424
        & 0.0405
        & 0.0371
        & \underline{0.0618}
        & 0.0610
        & \textbf{0.0620}
        & 0.0450
        & 0.0568
        & 0.0609
        & 0.0563
        & 0.0616
        \\

        & Hit@10
        & 0.0396
        & 0.0536
        & 0.0584
        & 0.0607
        & 0.0652
        & 0.0623
        & 0.0582
        & 0.0836
        & 0.0838
        & 0.0842
        & 0.0704
        & 0.0746
        & \textbf{0.0880}
        & 0.0791
        & \underline{0.0854}
        \\

        & NDCG@5
        & 0.0146
        & 0.0196
        & 0.0265
        & 0.0254
        & 0.0280
        & 0.0267
        & 0.0253
        & \textbf{0.0451}
        & 0.0428
        & \underline{0.0440}
        & 0.0262
        & 0.0401
        & 0.0415
        & 0.0398
        & \underline{0.0440}
        \\

        & NDCG@10
        & 0.0199
        & 0.0266
        & 0.0326
        & 0.0320
        & 0.0353
        & 0.0337
        & 0.0321
        & \textbf{0.0525}
        & 0.0502
        & 0.0512
        & 0.0344
        & 0.0467
        & 0.0504
        & 0.0471
        & \underline{0.0517}
        \\

        \midrule

        \multirow{4}{*}{\textbf{Sports}}
        & Hit@5
        & 0.0102
        & 0.0183
        & 0.0190
        & 0.0199
        & 0.0268
        & 0.0215
        & 0.0240
        & 0.0281
        & 0.0301
        & \underline{0.0311}
        & 0.0214
        & 0.0294
        & \textbf{0.0324}
        & 0.0288
        & \textbf{0.0324}
        \\

        & Hit@10
        & 0.0175
        & 0.0313
        & 0.0312
        & 0.0301
        & 0.0343
        & 0.0347
        & 0.0403
        & 0.0441
        & 0.0416
        & 0.0427
        & 0.0332
        & 0.0422
        & \textbf{0.0476}
        & 0.0412
        & \underline{0.0473}
        \\

        & NDCG@5
        & 0.0065
        & 0.0109
        & 0.0122
        & 0.0106
        & 0.0173
        & 0.0137
        & 0.0156
        & 0.0184
        & 0.0212
        & \underline{0.0218}
        & 0.0116
        & 0.0203
        & 0.0214
        & 0.0199
        & \textbf{0.0226}
        \\

        & NDCG@10
        & 0.0088
        & 0.0150
        & 0.0161
        & 0.0141
        & 0.0226
        & 0.0179
        & 0.0209
        & 0.0236
        & 0.0250
        & 0.0256
        & 0.0154
        & 0.0245
        & \underline{0.0257}
        & 0.0239
        & \textbf{0.0274}
        \\

        \midrule

        \multirow{4}{*}{\textbf{Toys}}
        & Hit@5
        & 0.0215
        & 0.0326
        & 0.0330
        & 0.0448
        & 0.0366
        & 0.0337
        & 0.0321
        & 0.0584
        & 0.0623
        & \textbf{0.0665}
        & 0.0523
        & 0.0615
        & 0.0594
        & 0.0579
        & \underline{0.0647}
        \\

        & Hit@10
        & 0.0332
        & 0.0517
        & 0.0490
        & 0.0626
        & 0.0566
        & 0.0547
        & 0.0512
        & 0.0714
        & 0.0843
        & \textbf{0.0901}
        & 0.0764
        & 0.0838
        & 0.0828
        & 0.0797
        & \underline{0.0877}
        \\

        & NDCG@5
        & 0.0131
        & 0.0192
        & 0.0228
        & 0.0300
        & 0.0245
        & 0.0209
        & 0.0210
        & 0.0464
        & 0.0453
        & \textbf{0.0476}
        & 0.0298
        & 0.0434
        & 0.0425
        & 0.0412
        & \underline{0.0467}
        \\

        & NDCG@10
        & 0.0168
        & 0.0254
        & 0.0279
        & 0.0358
        & 0.0309
        & 0.0276
        & 0.0272
        & 0.0505
        & 0.0524
        & \textbf{0.0552}
        & 0.0376
        & 0.0509
        & 0.0505
        & 0.0482
        & \underline{0.0541}
        \\

        \bottomrule
    \end{tabular}%
    }
\end{table*}


\begin{table}[tp]
    \centering
    \caption{Direct and Reason scores. Reason denotes natural-language reasoning for OneRec-Think and SID Routing for CogRec.}
    \label{tab:cot_vs_direct}
    \begin{tabular}{@{}llrrrr@{}}
        \toprule
        \multirow{2}{*}{\textbf{Dataset}} &
        \multirow{2}{*}{\textbf{Method}} &
        \multicolumn{2}{c}{\textbf{Hit@10}} &
        \multicolumn{2}{c}{\textbf{NDCG@10}} \\
        \cmidrule(lr){3-4}\cmidrule(lr){5-6}
        & & Direct & Reason & Direct & Reason \\
        \midrule
        \multirow{2}{*}{Beauty}
            & OneRec-Think & 0.0838 & 0.0834 & 0.0502 & 0.0495 \\
            & CogRec       & 0.0842 & 0.0854 & 0.0512 & 0.0517 \\
        \midrule
        \multirow{2}{*}{Sports}
            & OneRec-Think & 0.0416 & 0.0392 & 0.0250 & 0.0226 \\
            & CogRec       & 0.0427 & 0.0473 & 0.0256 & 0.0274 \\
        \midrule
        \multirow{2}{*}{Toys}
            & OneRec-Think & 0.0843 & 0.0852 & 0.0524 & 0.0535 \\
            & CogRec       & 0.0901 & 0.0877 & 0.0552 & 0.0541 \\
        \bottomrule
    \end{tabular}
\end{table}

\subsection{Multi-Stage Training Pipeline}
\label{sec:training_pipeline}

CogRec adopts the pipeline shown in Figure~\ref{fig:cogrec_overview}. The stages progressively introduce the SID vocabulary, direct recommendation capability, and alternative reasoning forms while maintaining a shared SID output space.

\textbf{Stage 0: SID Topology Initialization.}
This offline stage constructs the four-layer SIDs, intra-layer graphs, item neighborhood index, and data splits. It extends the base-model vocabulary with 1,024 layer-specific SID tokens and two boundary tokens.

\textbf{Stage 1: SID Token Alignment.}
Newly introduced SID tokens are randomly initialized. Stage 1 trains their embeddings using item descriptions paired with corresponding SID sequences while freezing other parameters. The checkpoint with the lowest validation loss serves as the common initialization.

\textbf{Stage 2: Direct SID Generation.}
Stage 2 trains the model to predict the target SID directly from the user history. The assistant response contains an empty \texttt{<think>} segment followed by the ground-truth SID. This stage produces the shared checkpoint for both reasoning branches.

\textbf{Stage 3a: Natural-Language Reasoning.}
Starting from the Stage 2 checkpoint, this branch generates a category level or preference level rationale before the target SID. It provides a controlled baseline whose intermediate text is semantically descriptive but not explicitly grounded in the SID topology.

\textbf{Stage 3b: SID Routing Reasoning.}
Also starting from the Stage 2 checkpoint, this branch replaces natural-language rationales with structured routing traces. Each assistant response contains the non-trivial \textsc{LateralJump} and \textsc{Explore} operations followed by the target SID.

This pipeline yields three controlled output formats: direct SID generation from Stage 2, natural-language reasoning from Stage 3a, and structure-grounded SID Routing from Stage 3b. Their comparison focuses on whether the form and grounding of the intermediate trace affect final SID generation.

\subsection{Training Objective and Constrained Decoding}
\label{sec:training_decoding}

For a formatted token sequence $Y=[y_1,\ldots,y_n]$, all trainable stages minimize the autoregressive loss
\begin{equation}
    \mathcal{L}
    =-\sum_{m=1}^{n} \alpha_m \log p_\theta(y_m\mid y_{<m}),
\end{equation}
\noindent where $\alpha_m$ is a stage-dependent loss mask. Stage 1 computes the loss over the alignment sequence. For Stages 2, 3a, and 3b, system-prompt tokens are masked, whereas the user-history and assistant-response segments participate in training.

The user-history segment contains many newly introduced SID tokens. Applying the language-modeling loss to this segment supplies direct next-token supervision for historical SID tokens, in addition to the loss on the reasoning trace and target SID.

At inference time, the model receives only the user history. In the direct format, it generates the target SID immediately. In a reasoning-augmented format, it first generates the intermediate segment and then the target SID. The target-SID portion is decoded with trie-constrained beam search, where the trie contains all valid item SID sequences. At every decoding step, only legal continuations of the current prefix are permitted. Consequently, all generative variants return identifiers corresponding to existing items and share the same valid output space.

\section{Experiments}
\label{sec:experiments}
In this section, we conduct experiments to evaluate CogRec. We address the following research questions:
\begin{itemize}
    \item \textbf{RQ1}: How does CogRec compare with state-of-the-art baselines, including traditional sequential recommendation and generative recommendation with or without reasoning enhancement?
    \item \textbf{RQ2}: How does the intermediate reasoning format affect recommendation performance when comparing direct SID generation, natural-language reasoning, and structure-grounded SID Routing?
    \item \textbf{RQ3}: How does SID Routing perform across structurally defined difficulty groups?
    \item \textbf{RQ4}: What structural route complexity and generated reasoning lengths are observed?
    \item \textbf{RQ5}: How does performance respond to sequential modifications of key design components, including the routing trace, loss mask, and learning-rate schedule?
\end{itemize}

\subsection{Experimental Setup}
\label{sec:exp_setup}

\subsubsection{Datasets}

We conduct experiments on three categories from the Amazon Reviews 2014 benchmark: Beauty, Sports, and Toys. Following common sequential recommendation protocols~\cite{kang2018sasrec,sun2019bert4rec,rajput2023tiger}, user interactions are sorted chronologically. For each user, the last interacted item is used for testing, the second-to-last item is used for validation, and the remaining interactions are used for training. Dataset statistics are summarized in Table~\ref{tab:dataset_stats}. Each item is represented by its title, description, and category, encoded into embeddings and quantized into four-layer Semantic IDs. The OneRec-Think and CogRec variants reproduced in our experimental setting share the same SID mapping.

\subsubsection{Evaluation Metrics}

Following standard evaluation protocols in sequential and Semantic-ID-based generative recommendation~\cite{kang2018sasrec,sun2019bert4rec,rajput2023tiger}, we use Hit@$K$ and NDCG@$K$ as evaluation metrics. Hit@$K$ checks whether the ground-truth item appears among the top-$K$ predictions, whereas NDCG@$K$ additionally discounts a correct prediction according to its ranking position. All generative outputs are mapped to valid candidate items via trie-constrained decoding before metric computation.

 \begin{table}[tp]
    \centering
    \caption{Dataset statistics.}
    \label{tab:dataset_stats}
    \begin{tabular}{lrrrr}
        \toprule
        Dataset & \#Users & \#Items & \#Interactions & Avg. Len. \\
        \midrule
        Beauty & 22,363 & 12,101 & 176,139 & 8.87 \\
        Sports & 35,598 & 18,357 & 260,739 & 8.32 \\
        Toys   & 19,412 & 11,924 & 148,185 & 8.63 \\
        \bottomrule
    \end{tabular}
\end{table}

\begin{figure*}[t]
    \centering
    \includegraphics[width=\textwidth]{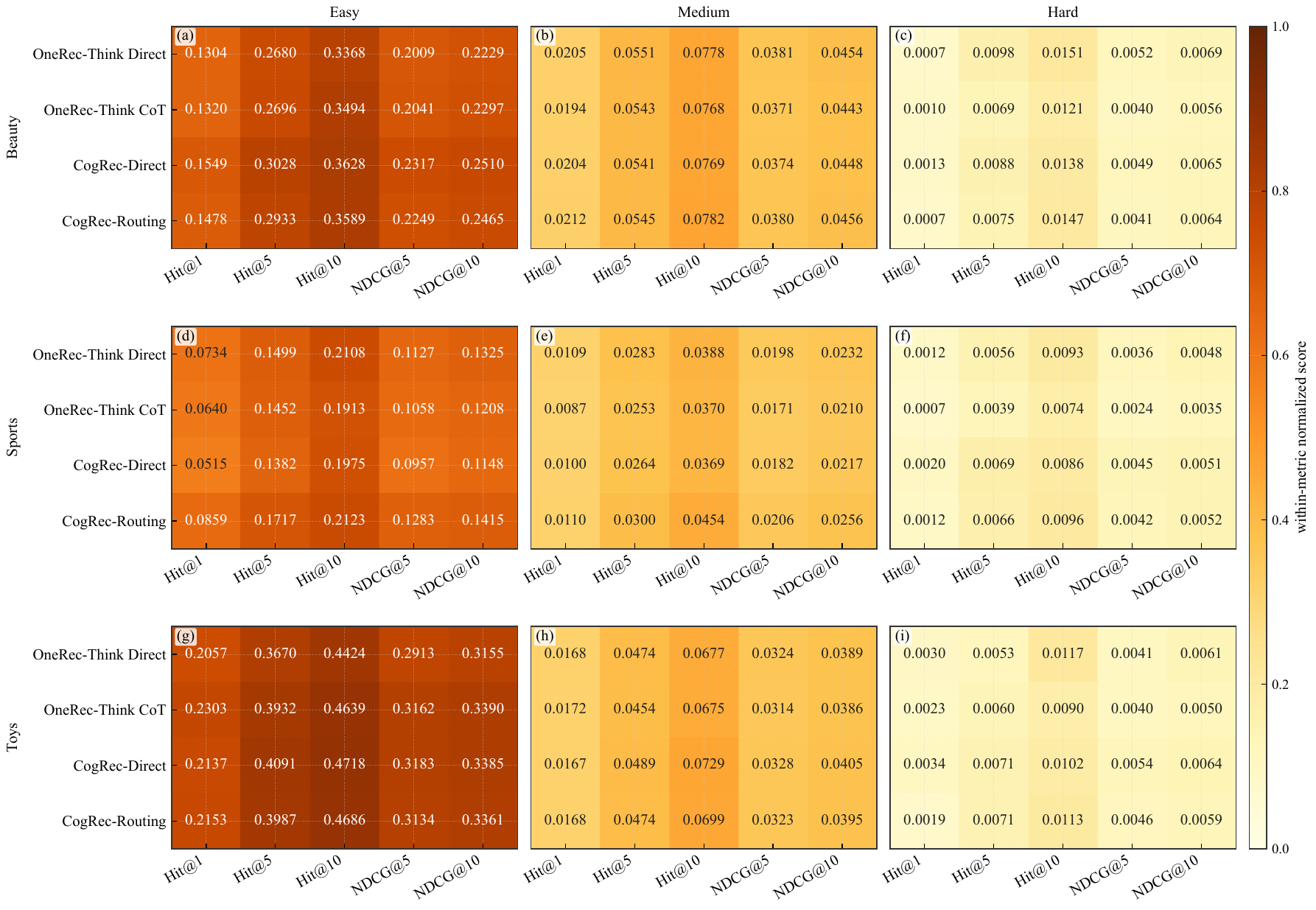}
    \caption{
    Difficulty-stratified performance of reproduced OneRec-Think and CogRec branches across dataset-difficulty groups. Columns report Hit@1, Hit@5, Hit@10, NDCG@5, and NDCG@10; Hit@1 is an auxiliary diagnostic available only under controlled evaluation.}
    \Description{
        Nine heat maps arranged by dataset and structural difficulty. Rows compare the reproduced OneRec-Think Direct and CoT branches with CogRec-Direct and CogRec-Routing. Columns report Hit@1, Hit@5, Hit@10, NDCG@5, and NDCG@10. Each cell contains the original score, and color intensity is normalized within the corresponding metric and panel.
        }
    \label{fig:difficulty_heatmap}
\end{figure*}

\subsubsection{Baselines}
We organize baselines into three groups.
\begin{itemize}
    \item \textbf{Traditional sequential recommendations}: GRU4Rec~\cite{hidasi2016gru4rec}, SASRec~\cite{kang2018sasrec}, BERT4Rec~\cite{sun2019bert4rec}, HGN~\cite{ma2019hgn}, and HSTU\footnote{\url{https://github.com/meta-recsys/generative-recommenders}}~\cite{zhai2024hstu}. These models operate on item identifiers or learned embeddings without explicit reasoning traces.
    
    \item \textbf{Generative recommendations}: TIGER~\cite{rajput2023tiger}, LETTER\footnote{\url{https://github.com/HonghuiBao2000/LETTER}}~\cite{wang2025letter}, EAGER\footnote{\url{https://github.com/yewzz/EAGER}}~\cite{zhang2025eager}, and CogRec-Direct. LETTER results are taken from the RAGR~\cite{zhang2026ragr} standalone environment. CogRec-Direct is our direct SID generation counterpart.
    
    \item \textbf{Reasoning-enhanced recommendations}: ReaRec\footnote{\url{https://github.com/TangJiakai/ReaRec}}~\cite{tang2025rearec}, PauseRec~\cite{zhang2026pauserec}, TwiSTAR~\cite{cao2026twistar}, OneRec-Think\footnote{\url{https://github.com/wangshy31/OneRec-Think}}~\cite{liu2025onerecthink}, and CogRec-Routing. 
\end{itemize}

\subsubsection{Implementation and Decoding Details}
\label{sec:implementation_details}

We use Qwen3-1.7B~\cite{yang2025qwen3} as the backbone and BAAI/bge-base-en-v1.5~\cite{xiao2024cpack} as the item text encoder. Item metadata is encoded into 768-dimensional embeddings and quantized into four-layer Semantic IDs with codebook size 256 and 30 K-means iterations per layer using residual quantization~\cite{oord2017vqvae,jegou2011product}. For each centroid, we consider its top 16 neighbors in the same layer and retain those with cosine similarity of at least 0.15, as illustrated in the left panel of Figure~\ref{fig:sid_space_routing}. For item-level neighborhoods, we build an HNSW index~\cite{malkov2018hnsw} with $M=32$ and \texttt{efConstruction}=200 within each finest-layer cluster containing at least ten items.

Stage 1 trains newly introduced SID token embeddings at $1\times10^{-4}$ for up to 15 epochs while freezing the backbone. Stage 2 fine-tunes the full model for direct generation at $1\times10^{-5}$ with batch size 16 for 6 epochs using constant learning rate, warmup ratio 0.1, and weight decay 0.01. Stage 3a trains Natural-Language Reasoning and Stage 3b trains SID Routing from the Stage 2 checkpoint with the same settings for 3 epochs; checkpoints are fixed at epoch 2 for Beauty and Sports, and epoch 3 for Toys. All stages use AdamW ($\beta_1=0.9$, $\beta_2=0.999$, $\epsilon=10^{-8}$), seed 42, and DeepSpeed ZeRO-2 on NVIDIA A800 GPUs.

All generative models use trie-constrained decoding over valid SIDs. Direct generation outputs the target directly. For reasoning-augmented generation, we first sample five intermediate traces at temperature 1.5 and top-p 0.95. For each trace, we then decode the target SID with beam size 10, temperature 0.6, top-p 1.0, and a limit of eight tokens, yielding up to 50 hypotheses across the five traces. For CoT evaluation, user histories are truncated to the 95th percentile per dataset for memory management. The system prompt is masked from loss; user history and assistant responses participate in training.

\begin{figure}[tp]
    \centering
    \includegraphics[width=\linewidth]{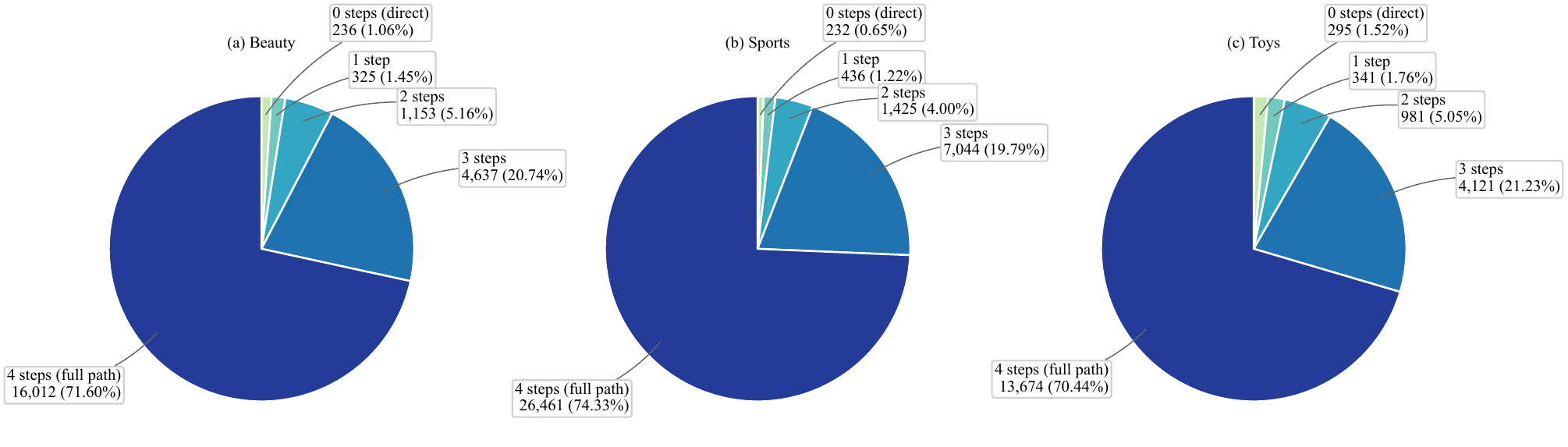}
    \caption{Distribution of non-\textsc{Match} operations in offline reference routes. One step denotes one layer with a non-\textsc{Match} operation.}
    \Description{
        Three pie charts for Beauty, Sports, and Toys. Each chart reports the percentage of test instances whose offline reference route contains zero, one, two, three, or four non-Match operations. In every dataset, the three-step and four-step categories form the large majority.
        }
    \label{fig:routing_step_distribution}
\end{figure}

\subsection{Overall Performance (RQ1)}
\label{sec:overall_performance}

Table~\ref{tab:overall_results} reports the performance of all compared methods across the three datasets. The table groups methods into traditional sequential recommenders, generative methods, and reasoning-enhanced methods. CogRec-Direct appears in the generative group because it predicts the target SID directly, while CogRec-Routing appears in the reasoning group because it generates a routing trace before the target SID. Baseline values follow their cited sources, including the LETTER results obtained from the RAGR standalone environment. This table therefore provides a cross-source comparison. The Direct-versus-Reason comparison within our experimental setting is reported separately in Table~\ref{tab:cot_vs_direct} and Figure~\ref{fig:difficulty_heatmap}. Among non-reasoning generative methods, CogRec-Direct achieves the strongest performance on Toys and remains competitive on Beauty and Sports, providing a strong direct-generation reference for assessing reasoning augmentation.

The effectiveness of reasoning enhancement is dataset-dependent. On Beauty, CogRec-Routing yields small gains over CogRec-Direct in Hit@10 and NDCG@10. On Sports, CogRec-Routing achieves the highest NDCG@5 and NDCG@10 among the compared methods. It also ties with TwiSTAR for the highest Hit@5 at the reported precision. On Toys, CogRec-Routing underperforms CogRec-Direct on all four reported metrics. These results show that the usefulness of an intermediate routing trace varies across datasets.

Across all datasets, CogRec-Routing is competitive with existing reasoning-enhanced recommendation approaches, including OneRec-Think and TwiSTAR. CogRec uses deterministically constructed, topology-derived traces for intermediate supervision, providing a distinct way to connect reasoning with the target SID space. The observed dataset-dependent variation motivates the reasoning-format comparison and difficulty-stratified analysis in the following sections.

\begin{figure}[tp]
    \centering
    \includegraphics[width=\linewidth]{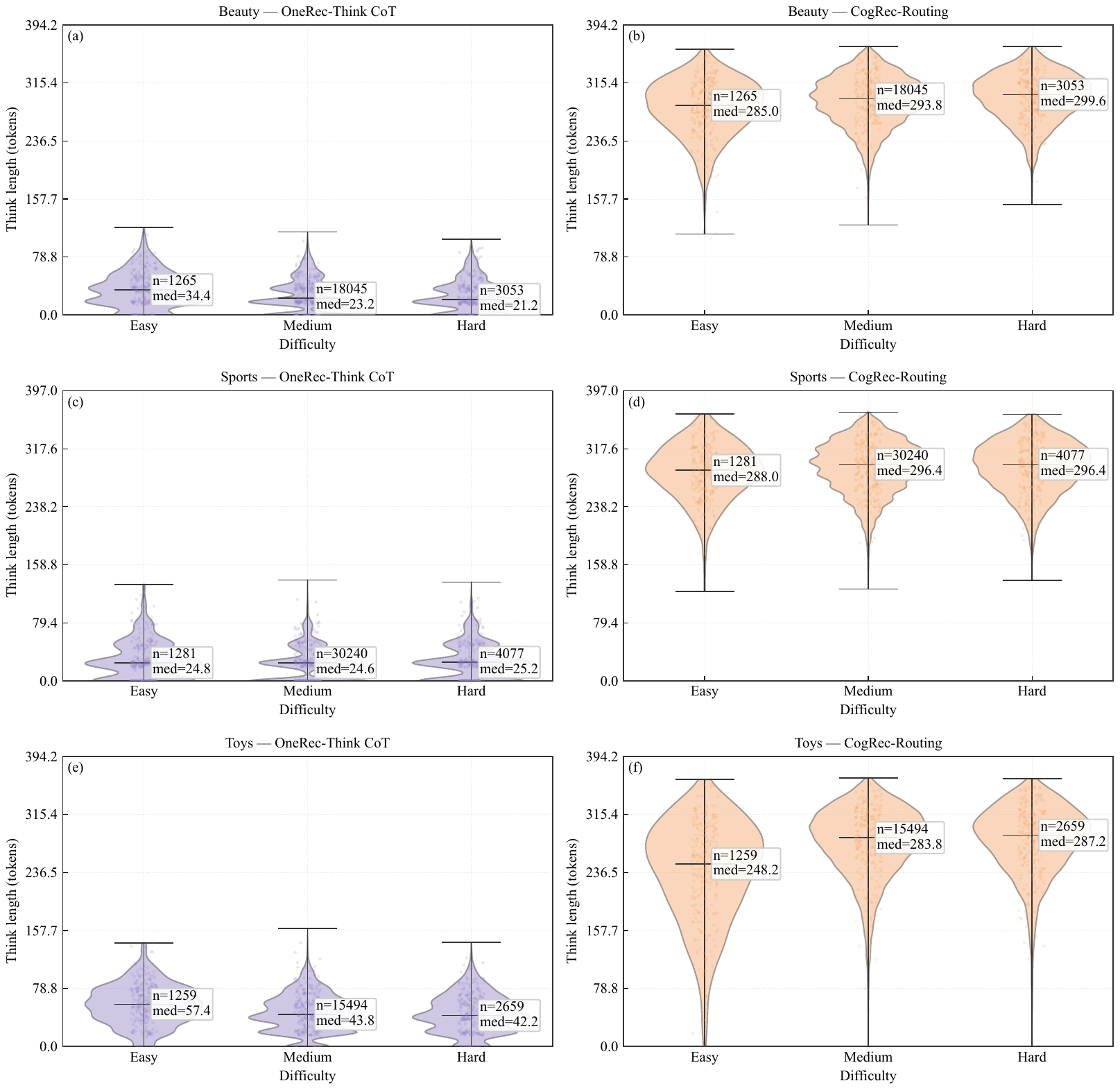}
    \caption{
        Think-length distribution for natural-language reasoning and SID Routing across structural difficulty groups. SID Routing produces longer intermediate traces.
    }
    \Description{
        Violin plots comparing the number of generated reasoning tokens for natural-language CoT and SID Routing across Easy, Medium, and Hard instances. SID Routing produces longer intermediate traces in all three difficulty groups.
    }
    \label{fig:think_length_distribution}
\end{figure}

\subsection{Effect of Reasoning Format (RQ2)}
\label{sec:reasoning_format}

RQ2 examines whether the representation of intermediate reasoning affects target SID generation. We compare direct generation, natural-language reasoning, and SID Routing using a shared SID mapping and the decoding settings described in Section~\ref{sec:implementation_details}. Table~\ref{tab:cot_vs_direct} reports the Direct and Reason scores side by side for OneRec-Think and CogRec on each dataset. Figure~\ref{fig:difficulty_heatmap} breaks down the performance of the same four branches by structural difficulty group.

Explicit reasoning does not universally improve recommendation. Direct generation remains highly competitive, especially on Toys. On Sports, CogRec-Routing scores higher than CogRec-Direct on both Hit@10 (0.0473 vs.\ 0.0427) and NDCG@10 (0.0274 vs.\ 0.0256). On the same dataset, OneRec-Think scores lower on both metrics with natural-language reasoning than with direct generation. These observations support a conditional view in which the effects of intermediate generation depend on the dataset and reasoning format. The routing-step distribution in Figure~\ref{fig:routing_step_distribution} shows that most offline reference routes contain at least three non-\textsc{Match} operations. The difficulty-stratified analysis that follows examines how performance varies across structurally defined groups.

\begin{table*}[tp]
    \centering
    \caption{
        Progressive design study on Beauty. Direct scores are evaluated on the full test set, while Reason scores are evaluated on a fixed 10\% test subset shared across all configurations.
    }
    \label{tab:design_sensitivity}
    \begin{tabular}{@{}cllcc@{}}
        \toprule
        \textbf{Step}
        & \textbf{Changed component}
        & \textbf{Cumulative configuration}
        & \makecell{\textbf{Direct H@10/N@10}\\\textbf{Full test set}}
        & \makecell{\textbf{Reason H@10/N@10}\\\textbf{10\% test subset}} \\
        \midrule
        0
        & Initial
        & Raw trace + User mask + Constant LR
        & 0.0842 / 0.0512
        & 0.0895 / 0.0538 \\

        1
        & Trace
        & Semantic-enriched + User mask + Constant LR
        & 0.0881 / 0.0529
        & 0.0367 / 0.0220 \\

        2
        & Loss mask
        & Semantic-enriched + Assistant-only + Constant LR
        & 0.0713 / 0.0414
        & 0.0327 / 0.0155 \\

        3
        & LR schedule
        & Semantic-enriched + Assistant-only + Cosine LR
        & 0.0805 / 0.0481
        & 0.0403 / 0.0215 \\
        \bottomrule
    \end{tabular}
\end{table*}

\subsection{Difficulty-Stratified Analysis (RQ3)}
\label{sec:difficulty_analysis}
RQ3 examines whether the benefit of SID Routing depends on the structural relation between the user history and the target. We partition test instances into Easy, Medium, and Hard groups using the prefix-overlap and graph-distance criteria defined in Section~\ref{sec:difficulty_step}. Group membership depends only on the history and the held-out target, so each test instance receives the same label across all four branches. Each panel in Figure~\ref{fig:difficulty_heatmap} compares the four branches for one dataset and difficulty group. Color intensity is normalized separately for each metric within a panel, while the printed scores allow comparisons across panels.

The results reveal variation across difficulty groups and datasets. On Easy instances, direct generation already performs strongly, while the gains from routing vary across datasets. On Medium instances, CogRec-Routing scores higher than CogRec-Direct on all five reported metrics for both Beauty and Sports. On Toys, only Hit@1 is higher, while Hit@5, Hit@10, NDCG@5, and NDCG@10 are all lower. Medium difficulty alone therefore does not identify when routing is beneficial. On Hard instances, absolute performance remains low across all methods, indicating that these structurally distant cases remain challenging. These findings support a conditional view of SID Routing.

\subsection{Routing Complexity and Reasoning Length (RQ4)}
\label{sec:routing_behavior}

RQ4 examines the structural complexity of reference routes and the length of generated reasoning traces. Figure~\ref{fig:routing_step_distribution} reports the number of non-\textsc{Match} operations per reference route, while Figure~\ref{fig:think_length_distribution} reports the token length of generated intermediate reasoning segments. A four-layer reference route contributes at most four steps, one for each non-\textsc{Match} layer. Generated length counts tokens in the intermediate text, so describing a single operation can contribute several tokens to a SID Routing trace.

The routing-step distribution reveals that reference routes containing three or four non-\textsc{Match} operations dominate all datasets, accounting for over 91\% of test instances. These routes encode non-matching relations at several SID layers, illustrating the structural content of the routing supervision. SID Routing also produces longer intermediate traces, reflecting the additional token generation used to express these relations.

Table~\ref{tab:quantitative_discussion} reports the share of Medium-difficulty instances and the share of three- or four-step reference routes alongside changes in Hit@10 and NDCG@10. For natural-language reasoning, each change is measured against OneRec-Think Direct; for SID Routing, it is measured against CogRec-Direct.

\begin{table}[tp]
    \centering
    \small
    \setlength{\tabcolsep}{2.3pt}
    \renewcommand{\arraystretch}{1.08}
    \caption{
    Quantitative summary of the reasoning regime. Medium denotes the proportion of Medium-difficulty instances; 3--4 steps denotes the proportion of routes with three or four non-Match operations. $\Delta$ is computed as reasoning-augmented minus direct-generation performance.
    }
    \label{tab:quantitative_discussion}
    \begin{tabular}{@{}lccrrrr@{}}
        \toprule
        \textbf{Dataset}
        & \makecell{\textbf{Medium}\\\textbf{(\%)}}
        & \makecell{\textbf{3--4}\\\textbf{steps (\%)}}
        & \multicolumn{2}{c}{
            \makecell{\textbf{Natural-language}\\\textbf{CoT $-$ Direct}}
        }
        & \multicolumn{2}{c}{
            \makecell{\textbf{SID Routing}\\\textbf{$-$ Direct}}
        } \\
        \cmidrule(lr){4-5}
        \cmidrule(lr){6-7}
        &
        &
        &
        $\Delta$H@10
        &
        $\Delta$N@10
        &
        $\Delta$H@10
        &
        $\Delta$N@10 \\
        \midrule

        Beauty
        & 80.69
        & 92.34
        & $-0.0004$
        & $-0.0007$
        & $+0.0012$
        & $+0.0005$ \\

        Sports
        & 84.95
        & 94.12
        & $-0.0024$
        & $-0.0024$
        & $+0.0046$
        & $+0.0018$ \\

        Toys
        & 79.82
        & 91.67
        & $+0.0009$
        & $+0.0011$
        & $-0.0024$
        & $-0.0011$ \\

        \bottomrule
    \end{tabular}
\end{table}

\subsection{Progressive Design Study (RQ5)}
\label{sec:design_sensitivity}

RQ5 evaluates a cumulative sequence of design changes on Beauty. To reduce the cost of reasoning evaluation, we construct a fixed subset-CoT set by uniformly sampling approximately 10\% of the test instances without replacement using random seed 42. All Reason configurations are evaluated on this same subset, while all Direct configurations are evaluated on the full test set. The full-test reasoning-format comparison is reported in Table~\ref{tab:cot_vs_direct}. Starting from Step 0 as the initial configuration, each subsequent step inherits all previous settings and changes one component: Step 1 replaces the raw routing trace with a semantic-enriched trace; Step 2 switches the loss mask from user-to-assistant to assistant-only supervision; Step 3 replaces the constant learning rate with cosine decay under the same semantic-enriched and assistant-only configuration. Results are presented in Table~\ref{tab:design_sensitivity}.

Step 1 shows higher Direct scores but sharply lower Reason scores under the semantic-enriched configuration. With assistant-only supervision at Step~2, both Direct and Reason scores fall relative to Step~1. Step 3 partially recovers both formats after switching to cosine decay, though performance remains below the initial configuration.

This trajectory describes a progressive failure-and-recovery pattern rather than isolated causal effects. Step~1 attains the highest Direct scores on the full Beauty test set, while Step~0 attains the highest Reason scores on the fixed subset. We retain Step~0 for its reasoning performance and use its compact raw routing trace, user-to-assistant supervision, and constant learning rate in the main experiments.

\subsection{Discussion}
The results suggest that the value of intermediate reasoning depends on the dataset and structural context. SID Routing connects intermediate generation to the output space by expressing item relations as layer-wise operations. This perspective emphasizes the alignment of reasoning supervision with target generation, while the variation across datasets and difficulty groups motivates examining when such structural guidance is useful.

\section{Conclusion}
We presented CogRec, a structure-cognitive framework for generative recommendation that uses SID topology as a source of intermediate reasoning supervision. It combines hierarchical identifiers with intra-layer semantic graphs and learns direct generation and structured SID Routing through a multi-stage pipeline. The study provides a concrete basis for analyzing structure-grounded recommendation reasoning. Future work will explore richer item relations and domain knowledge for routing supervision, along with more efficient decoding strategies.


\section{Ethical Considerations}
This work studies generative recommendation on public benchmark datasets and does not use private user data. Like other recommender systems, CogRec may inherit popularity bias, exposure bias, or metadata bias from the training data. In addition, SID Routing traces should be interpreted as model-side structural explanations derived from the SID topology, rather than as faithful descriptions of a user's actual decision process. If deployed in real systems, appropriate safeguards are needed for privacy protection, user controllability, transparency, and bias monitoring.

\begin{acks}
This work was supported in part by CCF-Tencent under Grant No.~RAGR20260127.
\end{acks}

\bibliographystyle{ACM-Reference-Format}
\bibliography{references}

@inproceedings{tay2022dsi,
  title     = {Transformer Memory as a Differentiable Search Index},
  author    = {Tay, Yi and Tran, Vinh Q. and Dehghani, Mostafa and Ni, Jianmo and Bahri, Dara and Mehta, Harsh and Qin, Zhen and Hui, Kai and Zhao, Zhe and Gupta, Jai and Schuster, Tal and Cohen, William W. and Metzler, Donald},
  booktitle = {Advances in Neural Information Processing Systems},
  volume    = {35},
  pages     = {21831--21843},
  publisher = {Curran Associates, Inc.},
  address   = {Red Hook, NY, USA},
  year      = {2022}
}

@inproceedings{rajput2023tiger,
  title     = {Recommender Systems with Generative Retrieval},
  author    = {Rajput, Shashank and Mehta, Nikhil and Singh, Anima and Keshavan, Raghunandan H. and Vu, Trung and Heldt, Lukasz and Hong, Lichan and Tay, Yi and Tran, Vinh Q. and Samost, Jonah and Kula, Maciej and Chi, Ed H. and Sathiamoorthy, Maheswaran},
  booktitle = {Advances in Neural Information Processing Systems},
  volume    = {36},
  pages     = {10299--10315},
  publisher = {Curran Associates, Inc.},
  address   = {Red Hook, NY, USA},
  year      = {2023}
}

@misc{hidasi2016gru4rec,
  title        = {Session-based Recommendations with Recurrent Neural Networks},
  author       = {Hidasi, Bal{\'a}zs and Karatzoglou, Alexandros and Baltrunas, Linas and Tikk, Domonkos},
  howpublished = {International Conference on Learning Representations},
  year         = {2016},
  note         = {Published as a conference paper at ICLR 2016},
  url          = {https://openreview.net/forum?id=yoffK5KZSgQ}
}

@inproceedings{kang2018sasrec,
  title     = {Self-Attentive Sequential Recommendation},
  author    = {Kang, Wang-Cheng and McAuley, Julian},
  booktitle = {2018 IEEE International Conference on Data Mining},
  pages     = {197--206},
  publisher = {IEEE},
  address   = {Piscataway, NJ, USA},
  year      = {2018},
  doi       = {10.1109/ICDM.2018.00035}
}

@inproceedings{sun2019bert4rec,
  title     = {{BERT4Rec}: Sequential Recommendation with Bidirectional Encoder Representations from Transformer},
  author    = {Sun, Fei and Liu, Jun and Wu, Jian and Pei, Changhua and Lin, Xiao and Ou, Wenwu and Jiang, Peng},
  booktitle = {Proceedings of the 28th ACM International Conference on Information and Knowledge Management},
  pages     = {1441--1450},
  publisher = {Association for Computing Machinery},
  address   = {New York, NY, USA},
  year      = {2019},
  doi       = {10.1145/3357384.3357895},
  url       = {https://doi.org/10.1145/3357384.3357895}
}

@inproceedings{ma2019hgn,
  title     = {Hierarchical Gating Networks for Sequential Recommendation},
  author    = {Ma, Chen and Kang, Peng and Liu, Xue},
  booktitle = {Proceedings of the 25th ACM SIGKDD International Conference on Knowledge Discovery and Data Mining},
  pages     = {825--833},
  publisher = {Association for Computing Machinery},
  address   = {New York, NY, USA},
  year      = {2019},
  doi       = {10.1145/3292500.3330984},
  url       = {https://doi.org/10.1145/3292500.3330984}
}

@inproceedings{zhai2024hstu,
  title     = {Actions Speak Louder than Words: Trillion-Parameter Sequential Transducers for Generative Recommendations},
  author    = {Zhai, Jiaqi and Liao, Lucy and Liu, Xing and Wang, Yueming and Li, Rui and Cao, Xuan and Gao, Leon and Gong, Zhaojie and Gu, Fangda and He, Jiayuan and Lu, Yinghai and Shi, Yu},
  booktitle = {Proceedings of the 41st International Conference on Machine Learning},
  series    = {Proceedings of Machine Learning Research},
  volume    = {235},
  pages     = {58484--58509},
  publisher = {PMLR},
  address   = {Vienna, Austria},
  year      = {2024},
  url       = {https://proceedings.mlr.press/v235/zhai24a.html}
}

@misc{liu2025onerec,
  title         = {{OneRec}: Unifying Retrieve and Rank with Generative Recommender and Iterative Preference Alignment},
  author        = {Deng, Jiaxin and Wang, Shiyao and Cai, Kuo and Ren, Lejian and Hu, Qigen and Ding, Weifeng and Luo, Qiang and Zhou, Guorui},
  year          = {2025},
  eprint        = {2502.18965},
  archivePrefix = {arXiv},
  primaryClass  = {cs.IR},
  url           = {https://arxiv.org/abs/2502.18965}
}

@inproceedings{liu2025onerecthink,
  title     = {{O}ne{R}ec-Think: In-Text Reasoning for Generative Recommendation},
  author    = {Liu, Zhanyu and Wang, Shiyao and Wang, Xingmei and Zhang, Rongzhou and Deng, Jiaxin and Bao, Honghui and Zhang, Jinghao and Li, Wuchao and Zheng, PengFei and Wu, Xiangyu and Hu, Yifei and Hu, Qigen and Luo, Xinchen and Ren, Lejian and Zixing, Zhang and Wang, Qianqian and Cai, Kuo and Wu, Yunfan and Cheng, Hongtao and Cheng, Zexuan and Ren, Lu and Wang, Huanjie and Su, Yi and Tang, Ruiming and Gai, Kun and Zhou, Guorui},
  editor    = {Liakata, Maria and Moreira, Viviane P. and Zhang, Jiajun and Jurgens, David},
  booktitle = {Proceedings of the 64th Annual Meeting of the Association for Computational Linguistics (Volume 1: Long Papers)},
  month     = jul,
  year      = {2026},
  address   = {San Diego, California, United States},
  publisher = {Association for Computational Linguistics},
  pages     = {2664--2681},
  isbn      = {979-8-89176-390-6},
  doi       = {10.18653/v1/2026.acl-long.123},
  url       = {https://aclanthology.org/2026.acl-long.123/}
}

@inproceedings{wang2025letter,
  title     = {Learnable Item Tokenization for Generative Recommendation},
  author    = {Wang, Wenjie and Bao, Honghui and Lin, Xinyu and Zhang, Jizhi and Li, Yongqi and Feng, Fuli and Ng, See-Kiong and Chua, Tat-Seng},
  booktitle = {Proceedings of the 33rd ACM International Conference on Information and Knowledge Management},
  pages     = {2400--2409},
  publisher = {Association for Computing Machinery},
  address   = {New York, NY, USA},
  year      = {2024},
  doi       = {10.1145/3627673.3679569},
  url       = {https://doi.org/10.1145/3627673.3679569}
}

@inproceedings{zhang2025eager,
  title     = {{EAGER}: Two-Stream Generative Recommender with Behavior-Semantic Collaboration},
  author    = {Wang, Ye and Xun, Jiahao and Hong, Minjie and Zhu, Jieming and Jin, Tao and Lin, Wang and Li, Haoyuan and Li, Linjun and Xia, Yan and Zhao, Zhou and Dong, Zhenhua},
  booktitle = {Proceedings of the 30th ACM SIGKDD Conference on Knowledge Discovery and Data Mining},
  pages     = {3245--3254},
  publisher = {Association for Computing Machinery},
  address   = {New York, NY, USA},
  year      = {2024},
  doi       = {10.1145/3637528.3671775},
  url       = {https://doi.org/10.1145/3637528.3671775}
}

@misc{zhang2026pauserec,
  title         = {Implicit Reasoning for Large Language Model-based Generative Recommendation},
  author        = {He, Yinhan and Collins, Liam and Kumar, Bhuvesh and Li, Jundong and Shah, Neil and Loveland, Donald},
  year          = {2026},
  eprint        = {2606.14142},
  archivePrefix = {arXiv},
  primaryClass  = {cs.CL},
  url           = {https://arxiv.org/abs/2606.14142}
}

@misc{cao2026twistar,
  title         = {{TwiSTAR}: Think Fast, Think Slow, Then Act, Generative Recommendation with Adaptive Reasoning},
  author        = {Cao, Shiteng and Jiang, Kaian and Gong, Yunlong and Li, Zhiheng},
  year          = {2026},
  eprint        = {2605.11553},
  archivePrefix = {arXiv},
  primaryClass  = {cs.IR},
  url           = {https://arxiv.org/abs/2605.11553}
}

@inproceedings{geng2022p5,
  title     = {Recommendation as Language Processing ({RLP}): A Unified Pretrain, Personalized Prompt and Predict Paradigm ({P5})},
  author    = {Geng, Shijie and Liu, Shuchang and Fu, Zuohui and Ge, Yingqiang and Zhang, Yongfeng},
  booktitle = {Proceedings of the 16th ACM Conference on Recommender Systems},
  pages     = {299--315},
  publisher = {Association for Computing Machinery},
  address   = {New York, NY, USA},
  year      = {2022},
  doi       = {10.1145/3523227.3546767},
  url       = {https://doi.org/10.1145/3523227.3546767}
}

@article{wu2024llmrecsurvey,
  title     = {A Survey on Large Language Models for Recommendation},
  author    = {Wu, Likang and Zheng, Zhi and Qiu, Zhaopeng and Wang, Hao and Gu, Hongchao and Shen, Tingjia and Qin, Chuan and Zhu, Chen and Zhu, Hengshu and Liu, Qi and Xiong, Hui and Chen, Enhong},
  journal   = {World Wide Web},
  volume    = {27},
  number    = {5},
  articleno = {60},
  numpages  = {31},
  publisher = {Springer},
  year      = {2024},
  doi       = {10.1007/s11280-024-01291-2},
  url       = {https://doi.org/10.1007/s11280-024-01291-2}
}

@inproceedings{wei2022cot,
  title     = {Chain-of-Thought Prompting Elicits Reasoning in Large Language Models},
  author    = {Wei, Jason and Wang, Xuezhi and Schuurmans, Dale and Bosma, Maarten and Ichter, Brian and Xia, Fei and Chi, Ed H. and Le, Quoc V. and Zhou, Denny},
  booktitle = {Advances in Neural Information Processing Systems},
  volume    = {35},
  pages     = {24824--24837},
  publisher = {Curran Associates, Inc.},
  address   = {Red Hook, NY, USA},
  year      = {2022}
}

@misc{yao2023react,
  title        = {{ReAct}: Synergizing Reasoning and Acting in Language Models},
  author       = {Yao, Shunyu and Zhao, Jeffrey and Yu, Dian and Du, Nan and Shafran, Izhak and Narasimhan, Karthik and Cao, Yuan},
  howpublished = {The Eleventh International Conference on Learning Representations},
  year         = {2023},
  note         = {Published as a conference paper at ICLR 2023},
  url          = {https://openreview.net/forum?id=WE_vluYUL-X}
}

@inproceedings{pan2025dynathink,
  title     = {{DynaThink}: Fast or Slow? A Dynamic Decision-Making Framework for Large Language Models},
  author    = {Pan, Jiabao and Zhang, Yan and Zhang, Chen and Liu, Zuozhu and Wang, Hongwei and Li, Haizhou},
  booktitle = {Proceedings of the 2024 Conference on Empirical Methods in Natural Language Processing},
  pages     = {14686--14695},
  publisher = {Association for Computational Linguistics},
  address   = {Miami, Florida, USA},
  year      = {2024},
  doi       = {10.18653/v1/2024.emnlp-main.814},
  url       = {https://aclanthology.org/2024.emnlp-main.814/}
}

@inproceedings{chung2025thinker,
  title     = {Thinker: Learning to Think Fast and Slow},
  author    = {Chung, Stephen and Du, Wenyu and Fu, Jie},
  booktitle = {Advances in Neural Information Processing Systems},
  volume    = {38},
  pages     = {102812--102841},
  publisher = {Curran Associates, Inc.},
  address   = {Red Hook, NY, USA},
  year      = {2025},
  url       = {https://proceedings.neurips.cc/paper_files/paper/2025/hash/94dc604e115237a7f4a758b3146cd976-Abstract-Conference.html}
}

@misc{tang2025rearec,
  title        = {Think Before Recommend: Unleashing the Latent Reasoning Power for Sequential Recommendation},
  author       = {Tang, Jiakai and Dai, Sunhao and Shi, Teng and Xu, Jun and Chen, Xu and Chen, Wen and Wu, Jian and Jiang, Yuning},
  howpublished = {IEEE Transactions on Knowledge and Data Engineering, Early Access},
  year         = {2026},
  doi          = {10.1109/TKDE.2026.3694421},
  url          = {https://doi.org/10.1109/TKDE.2026.3694421}
}

@book{kahneman2011thinking,
  title     = {Thinking, Fast and Slow},
  author    = {Kahneman, Daniel},
  publisher = {Farrar, Straus and Giroux},
  address   = {New York, NY, USA},
  year      = {2011}
}

@inproceedings{wang2019ngcf,
  title     = {Neural Graph Collaborative Filtering},
  author    = {Wang, Xiang and He, Xiangnan and Wang, Meng and Feng, Fuli and Chua, Tat-Seng},
  booktitle = {Proceedings of the 42nd International ACM SIGIR Conference on Research and Development in Information Retrieval},
  pages     = {165--174},
  publisher = {Association for Computing Machinery},
  address   = {New York, NY, USA},
  year      = {2019},
  doi       = {10.1145/3331184.3331267},
  url       = {https://doi.org/10.1145/3331184.3331267}
}

@inproceedings{he2020lightgcn,
  title     = {{LightGCN}: Simplifying and Powering Graph Convolution Network for Recommendation},
  author    = {He, Xiangnan and Deng, Kuan and Wang, Xiang and Li, Yan and Zhang, Yongdong and Wang, Meng},
  booktitle = {Proceedings of the 43rd International ACM SIGIR Conference on Research and Development in Information Retrieval},
  pages     = {639--648},
  publisher = {Association for Computing Machinery},
  address   = {New York, NY, USA},
  year      = {2020},
  doi       = {10.1145/3397271.3401063},
  url       = {https://doi.org/10.1145/3397271.3401063}
}

@inproceedings{oord2017vqvae,
  title     = {Neural Discrete Representation Learning},
  author    = {van den Oord, Aaron and Vinyals, Oriol and Kavukcuoglu, Koray},
  booktitle = {Advances in Neural Information Processing Systems},
  volume    = {30},
  pages     = {6306--6315},
  publisher = {Curran Associates, Inc.},
  address   = {Red Hook, NY, USA},
  year      = {2017}
}

@article{jegou2011product,
  title     = {Product Quantization for Nearest Neighbor Search},
  author    = {J{\'e}gou, Herv{\'e} and Douze, Matthijs and Schmid, Cordelia},
  journal   = {IEEE Transactions on Pattern Analysis and Machine Intelligence},
  volume    = {33},
  number    = {1},
  pages     = {117--128},
  publisher = {IEEE},
  year      = {2011},
  doi       = {10.1109/TPAMI.2010.57},
  url       = {https://doi.org/10.1109/TPAMI.2010.57}
}

@article{malkov2018hnsw,
  title     = {Efficient and Robust Approximate Nearest Neighbor Search Using Hierarchical Navigable Small World Graphs},
  author    = {Malkov, Yu. A. and Yashunin, D. A.},
  journal   = {IEEE Transactions on Pattern Analysis and Machine Intelligence},
  volume    = {42},
  number    = {4},
  pages     = {824--836},
  publisher = {IEEE},
  year      = {2020},
  doi       = {10.1109/TPAMI.2018.2889473},
  url       = {https://doi.org/10.1109/TPAMI.2018.2889473}
}

@inproceedings{xiao2024cpack,
  title     = {{C-Pack}: Packed Resources for General Chinese Embeddings},
  author    = {Xiao, Shitao and Liu, Zheng and Zhang, Peitian and Muennighoff, Niklas and Lian, Defu and Nie, Jian-Yun},
  booktitle = {Proceedings of the 47th International ACM SIGIR Conference on Research and Development in Information Retrieval},
  pages     = {641--649},
  publisher = {Association for Computing Machinery},
  address   = {New York, NY, USA},
  year      = {2024},
  doi       = {10.1145/3626772.3657878},
  url       = {https://doi.org/10.1145/3626772.3657878}
}

@misc{yang2025qwen3,
  title         = {{Qwen3} Technical Report},
  author        = {Yang, An and Li, Anfeng and Yang, Baosong and Zhang, Beichen and Hui, Binyuan and Zheng, Bo and Yu, Bowen and Gao, Chang and Huang, Chengen and Lv, Chenxu and Zheng, Chujie and Liu, Dayiheng and Zhou, Fan and Huang, Fei and Hu, Feng and Ge, Hao and Wei, Haoran and Lin, Huan and Tang, Jialong and Yang, Jian and Tu, Jianhong and Zhang, Jianwei and Yang, Jianxin and Yang, Jiaxi and Zhou, Jing and Zhou, Jingren and Lin, Junyang and Dang, Kai and Bao, Keqin and Yang, Kexin and Yu, Le and Deng, Lianghao and Li, Mei and Xue, Mingfeng and Li, Mingze and Zhang, Pei and Wang, Peng and Zhu, Qin and Men, Rui and Gao, Ruize and Liu, Shixuan and Luo, Shuang and Li, Tianhao and Tang, Tianyi and Yin, Wenbiao and Ren, Xingzhang and Wang, Xinyu and Zhang, Xinyu and Ren, Xuancheng and Fan, Yang and Su, Yang and Zhang, Yichang and Zhang, Yinger and Wan, Yu and Liu, Yuqiong and Wang, Zekun and Cui, Zeyu and Zhang, Zhenru and Zhou, Zhipeng and Qiu, Zihan},
  year          = {2025},
  eprint        = {2505.09388},
  archivePrefix = {arXiv},
  primaryClass  = {cs.CL},
  url           = {https://arxiv.org/abs/2505.09388}
}

@misc{wang2023selfconsistency,
  title        = {Self-Consistency Improves Chain of Thought Reasoning in Language Models},
  author       = {Wang, Xuezhi and Wei, Jason and Schuurmans, Dale and Le, Quoc V. and Chi, Ed H. and Narang, Sharan and Chowdhery, Aakanksha and Zhou, Denny},
  howpublished = {The Eleventh International Conference on Learning Representations},
  year         = {2023},
  note         = {Published as a conference paper at ICLR 2023},
  url          = {https://openreview.net/forum?id=1PL1NIMMrw}
}

@inproceedings{yao2023tot,
  title     = {Tree of Thoughts: Deliberate Problem Solving with Large Language Models},
  author    = {Yao, Shunyu and Yu, Dian and Zhao, Jeffrey and Shafran, Izhak and Griffiths, Thomas L. and Cao, Yuan and Narasimhan, Karthik},
  booktitle = {Advances in Neural Information Processing Systems},
  volume    = {36},
  pages     = {11809--11822},
  publisher = {Curran Associates, Inc.},
  address   = {Red Hook, NY, USA},
  year      = {2023},
  url       = {https://proceedings.neurips.cc/paper_files/paper/2023/hash/271db9922b8d1f4dd7aaef84ed5ac703-Abstract-Conference.html}
}

@misc{zhou2023leasttomost,
  title        = {Least-to-Most Prompting Enables Complex Reasoning in Large Language Models},
  author       = {Zhou, Denny and Sch{\"a}rli, Nathanael and Hou, Le and Wei, Jason and Scales, Nathan and Wang, Xuezhi and Schuurmans, Dale and Cui, Claire and Bousquet, Olivier and Le, Quoc V. and Chi, Ed H.},
  howpublished = {The Eleventh International Conference on Learning Representations},
  year         = {2023},
  note         = {Published as a conference paper at ICLR 2023},
  url          = {https://openreview.net/forum?id=WZH7099tgfM}
}

@inproceedings{lee2025gram,
  title     = {{GRAM}: Generative Recommendation via Semantic-aware Multi-granular Late Fusion},
  author    = {Lee, Sunkyung and Choi, Minjin and Choi, Eunseong and Kim, Hye-young and Lee, Jongwuk},
  booktitle = {Proceedings of the 63rd Annual Meeting of the Association for Computational Linguistics (Volume 1: Long Papers)},
  pages     = {33294--33312},
  address   = {Vienna, Austria},
  publisher = {Association for Computational Linguistics},
  year      = {2025},
  doi       = {10.18653/v1/2025.acl-long.1596},
  url       = {https://aclanthology.org/2025.acl-long.1596/}
}

@inproceedings{xie2025lohrec,
  title     = {{LOHRec}: Leveraging Order and Hierarchy in Generative Sequential Recommendation},
  author    = {Xie, Jiawen and Wu, Haiyang and Ji, Deyi and Yang, Yuekui and Ma, Shaoping},
  booktitle = {Findings of the Association for Computational Linguistics: EMNLP 2025},
  pages     = {17968--17983},
  address   = {Suzhou, China},
  publisher = {Association for Computational Linguistics},
  year      = {2025},
  doi       = {10.18653/v1/2025.findings-emnlp.977},
  url       = {https://aclanthology.org/2025.findings-emnlp.977/}
}

@inproceedings{fu2026diger,
  title     = {Differentiable Semantic ID for Generative Recommendation},
  author    = {Fu, Junchen and Ge, Xuri and Karatzoglou, Alexandros and Arapakis, Ioannis and Verberne, Suzan and Jose, Joemon M. and Ren, Zhaochun},
  booktitle = {Proceedings of the 49th International ACM SIGIR Conference on Research and Development in Information Retrieval},
  pages     = {369--379},
  publisher = {Association for Computing Machinery},
  address   = {New York, NY, USA},
  year      = {2026},
  doi       = {10.1145/3805712.3809641},
  url       = {https://doi.org/10.1145/3805712.3809641}
}

@misc{he2026sidreasoner,
  title         = {Reasoning over Semantic IDs Enhances Generative Recommendation},
  author        = {He, Yingzhi and Sun, Yan and Tan, Junfei and Chen, Yuxin and Kong, Xiaoyu and Shen, Chunxu and Wang, Xiang and Zhang, An and Chua, Tat-Seng},
  year          = {2026},
  eprint        = {2603.23183},
  archivePrefix = {arXiv},
  primaryClass  = {cs.IR},
  note          = {Accepted by the 32nd ACM SIGKDD Conference on Knowledge Discovery and Data Mining},
  url           = {https://arxiv.org/abs/2603.23183}
}

@misc{cheng2026capsid,
  title         = {{CapsID}: Soft-Routed Variable-Length Semantic IDs for Generative Recommendation},
  author        = {Cheng, Wenzhuo and Gong, Menghang and Guo, Qixin and Zheng, Hang and Yang, Zhaobin and Lou, Jianguo and Zheng, Zhengwei},
  year          = {2026},
  eprint        = {2605.05096},
  archivePrefix = {arXiv},
  primaryClass  = {cs.IR},
  url           = {https://arxiv.org/abs/2605.05096}
}

@misc{zhang2026ragr,
  title         = {{RAGR}: Review-Augmented Generative Recommendation},
  author        = {Zhang, Yingyi and Li, Junyi and Wang, Yejing and Zhang, Wenlin and Qian, Xiaowei and Zhang, Sheng and Feng, Yue and Wang, Yichao and Liu, Yong and Zhao, Xiangyu and Li, Xianneng},
  year          = {2026},
  eprint        = {2605.17267},
  archivePrefix = {arXiv},
  primaryClass  = {cs.IR},
  url           = {https://arxiv.org/abs/2605.17267}
}


\end{document}